\documentclass[12pt]{article}
\newcommand{\bb}{\begin{eqnarray}}
\newcommand{\ee}{\end{eqnarray}}
\usepackage{hyperref}
\begin{document}
\title{Understanding CP violation in lattice QCD}
\author{P. Mitra\thanks{mitra@tnp.saha.ernet.in}\\
Saha Institute of Nuclear Physics, \\
1/AF Bidhannagar,\\
Calcutta 700064, India}
\date{hep-lat/0102008}
\maketitle
\begin{abstract}
It is pointed out that any CP violation
which may be found in lattice QCD with
a chiral phase in the fermion mass term
but without an explicit theta term
cannot be relevant for the continuum theory.
CP is classically conserved in the corresponding
continuum theory and is non-anomalous.
\end{abstract}

\section{Introduction}
The strong interactions per se
do not violate parity, but the Lagrangian of QCD may contain a $\theta
{\rm tr}~F\tilde F$ term which violates parity and CP.   Further, the quark
mass term $\bar\psi me^{i\gamma_5\theta'}\psi$ has an unknown chiral phase
$\theta'$ which also appears to violate these.  However, there is no
experimental
evidence of such violations.  It is difficult to estimate directly the
effect of the gluonic $\theta$ term, but calculations and proposals are often
made with the $\theta'$ term, mainly in the continuum and sometimes on the
lattice.

This $\theta'$-term appears to violate CP classically \cite{baluni},
but it  is known \cite{bcm} that a classically and
perturbatively conserved parity can be defined in the presence of this term:
parity and CP violation can
then occur either if there is a $\theta$ term, which explicitly breaks parity,
or if induced by nontrivial topology of gauge fields in functional
integration, {\it i.e.,} if the symmetries develop nonperturbative anomalies
\cite{bcm,aokiplus,aokihat}.  A chiral transformation with an associated
anomaly does enter the definition of the conserved parity,
so it may not be surprising if the CP symmetry is afflicted with an anomaly
upon quantization. To understand whether the short distance
singularities of the quantum field theory have such an effect,
it is necessary to use a regularization.

A symmetry in a classical field theory can develop an anomaly in
the quantized theory if it is
not possible to find any regularization to preserve it. It is, of course, not
difficult to find regularizations which {\it violate} a given symmetry. To
prove that a classical symmetry is preserved in the quantum theory, it is
sufficient to find {\it one} regularization which preserves the symmetry,
whereas for the existence of an anomaly, it is necessary
that {\it all} regularizations break it.
This principle will be applied here to the question of parity or CP violation
in the presence of the $\theta'$ term.

A  generalized Pauli-Villars regularization has been used \cite{bcm2},
but it involves the introduction of unphysical regulator
species.  A lattice regularization can be used to avoid this complication and
{\it nonperturbative} anomalies can be sought to be detected
through the lattice.
Lattice calculations of CP violation in QCD have in fact been
considered in the literature \cite{aokiplus}. In this note we shall therefore
investigate this class of regularizations.
This will confirm that the classically conserved CP does not develop an
anomaly, so that  any CP violation observed in lattice
QCD with just a chiral phase $\theta'$ in the mass term
must be an artefact and will {\it not} be relevant for
continuum QCD. This means that not only quenched lattice calculations
with the $\theta'$ term but even dynamical calculations with it
\cite{aokiplus} have nothing to do with continuum QCD. The $\theta$ term
must be used for any significant result.

\section{Parity symmetry}
The spinor part of the single flavour continuum QCD euclidean action reads
\bb
S=\int d^4x\bar\psi[\gamma^\mu \Delta_\mu-me^{i\theta'\gamma_5}]\psi,
\label{S}\ee
where $\Delta_\mu$ stands for the covariant derivative.
When $\theta'$ is non-zero, this breaks the standard parity
transformation in which
$A_0$ transforms as a true scalar and $A_i$ as a polar vector
and the fermion field transforms as
\bb
\bar\psi(x_0,x_i)&\rightarrow &\bar\psi(x_0,-x_i)\gamma_0\nonumber\\
\psi(x_0,x_i)&\rightarrow & \gamma_0\psi(x_0,-x_i).
\ee
As mentioned above, there is a  parity transformation \cite{bcm} that
leaves $S$ invariant. This involves the usual transformation for gauge fields
but that
for fermions is altered to include a chiral rotation proportional to
$\theta'$:
\bb
\bar\psi(x_0,x_i)&\rightarrow &\bar\psi(x_0,-x_i)
e^{i\theta'\gamma_5}\gamma_0\nonumber\\
\psi(x_0,x_i)&\rightarrow &
\gamma_0e^{i\theta'\gamma_5}\psi(x_0,-x_i).
\label{parity}\ee
This $\theta'$-dependent parity is easily seen to leave the action, with
exactly the same chiral phase $\theta'$ in the mass term, invariant,
except for any $\theta$ term in the pure gauge piece of the action,
which of course changes sign.

This is not enough to guarantee that $\theta'$ does not violate CP
in the quantized theory, as
the altered parity transformation contains
a chiral transformation and
may consequently be anomalous.
To investigate this matter,
we examine the theory regularized by introduction of the lattice.

\subsection{Wilson regularization of fermions}
The standard lattice regularization for fermions with the Wilson prescription
is
\bb
S_{\rm W}=a^4\sum_x \bar\psi[\gamma^\mu {D_\mu+D^{*}_\mu\over 2}-
aD^{*}_\mu D_\mu -me^{i\theta'\gamma_5}]\psi.
\ee
Here, $a$ is the lattice spacing, $D_\mu$ the forward covariant difference
operator on the lattice and $D^{*}_\mu$ the backward covariant difference
operator.
The fermion action is understood to be supplemented with the gauge
field action which involves the standard plaquette contribution
together with a $\theta$ term, for which one may use one of the
available forms like \cite{vecchia}:
\bb
S_g=\sum_{plaquettes} S_{plaquette} + i\theta Q.
\ee

The fermionic part $S_{\rm W}$ of the action is {\it not} invariant under the
lattice version of (\ref{parity}), with the link variables transformed in the
usual way, and
one may think that the breaking of this classical symmetry of the unregularized
theory by a regularization is an anomaly. Unlike the
popular continuum approach to anomalies, this breaking does not arise from a
regularization of the measure, but the lattice provides a
regularization of the action, and the chiral anomaly is known to manifest
itself in the Wilson formulation of fermions
as an explicit breaking of the chiral symmetry of the action.

However, this is not the whole story. Before suspecting that the classical
parity symmetry  develops an anomaly, one has to check if this symmetry
is broken in {\it all} regularizations.
As recognized in \cite{seistam,creutz},
the $D^{*}D$ term in the lattice action can be assigned a chiral phase:
\bb
S^{\theta''}_{\rm W}=a^4\sum_x \bar\psi[\gamma^\mu {D_\mu+D^{*}_\mu\over 2}
-aD^{*}_\mu D_\mu
e^{i\theta''\gamma_5} -me^{i\theta'\gamma_5}]\psi.
\label{theta''}\ee
Thus within the Wilson formulation of lattice fermions, there
is a whole class of regularizations parametrized by $\theta''$.

\subsubsection{The symmetric choice}
Let us see what happens in the special case $\theta''=\theta'$.
In this case, the lattice
implementation of (\ref{parity}), with the link variables transformed in the
usual way, leaves the fermionic action (\ref{theta''}) invariant, {\it i.e.,}
this parity is not broken by the $\theta'$ term in the quantum theory.
This is  a very special,
non-generic regularization, but the existence of just one parity conserving
regularization means that the transformation (\ref{parity}) is not anomalous.
One can take the limit $a\to 0$ with $\theta''=\theta'$ and the parity
continues to be conserved in the continuum limit. The same is true of CP.
This is the result announced: CP is neither explicitly broken nor anomalous
if $\theta'\neq 0$.
Of course, the symmetry is explicitly broken if $\theta\neq 0$ in $S_g$.

\subsubsection{Other choices}
It may be of interest to see what happens
if one deliberately chooses a regularization with $\theta''\neq\theta'$.
It is straightforward to see that the term with the phase $\theta''$
is replaced, under (\ref{parity}), by a term with a phase
$2\theta'-\theta''$. As a regularization which preserves the symmetry
is available, this breaking by deliberate choice is an artefact
rather than an anomaly.
The explicit breaking in this class of regularizations
in the fermionic piece of the action
is governed by the effective CP violating parameter
\bb
\bar\theta_f=\theta'-\theta''.
\ee
In addition, as before, the $\theta$ term in $S_g$ also violates CP.
Thus there are two
CP violating parameters on the lattice, $\bar\theta_f$ and $\theta$.
They cannot be
transformed into each other through chiral rotations {\it on the lattice}
because the measure in this regularized theory is trivial.
To first order, one may split any measure of CP violation like the
electric dipole moment of the neutron in the form
\bb
\Delta (a)=\Delta_f(a)\bar\theta_f+\Delta_g(a)\theta,
\ee
where the $a$-dependent coefficients on the right hand side
may be different for the fermion and gauge sectors.
The behaviour of these two parameters in the continuum limit is of interest.
It is possible to make a general statement without detailed calculations.
In the limit $a\to 0$, where the regularization is removed,
a dependence on the regularization parameter $\theta''$
must not survive:
\bb
{\partial\Delta  (0)\over\partial\theta''}=0.
\ee
Without calculating $\Delta (a)$, one can then argue that
\bb
{\partial\bar\theta_f\over\partial\theta''}\Delta_f(0)=-\Delta_f(0)=0.
\ee
The fact that any CP violation from the quark sector
has to depend on $\theta'-\theta''$
thus requires the $\theta'$ dependence to vanish in the continuum.
The only continuum contribution that can survive is
\bb
\Delta  (0)=\Delta_g(0)\theta,
\ee
showing that only $\theta$ can be a physical parameter, $\theta'$ is not.
It is easy to generalize this to higher order effects.
Any CP violation in the continuum theory can consequently come only from
$\theta$.  The above arguments cannot indicate whether
the $\theta$ dependence is non-zero or not,
so one can say that the vacuum angle, if nonzero, {\it may} lead to
CP violation, but the phase in the fermion mass term
has no such effect.

This argument fits in with the result reached above
that the parity which is classically conserved in the presence
of $\theta'$ is not anomalous. The
regularization which is needed on the lattice to preserve
this symmetry has $\theta''=\theta'$, {\it i.e.,} $\bar\theta_f=0$.
The violation of the symmetry for any other value of this parameter
is a regularization artefact, because one symmetric
choice of the regularization parameter $\theta''$ is available.
It is well known that the breaking of
rotation invariance on the cubical lattice is not an
anomaly but a regularization artefact that
can be avoided by using random lattices or continuum regularizations
and in fact goes away in the continuum limit.  In
the same way, parity and CP are violated by $\theta'$
only in  regularizations with a different value of $\theta''$
and the violation does not survive in the continuum limit.

\subsection{Ginsparg-Wilson formulation}
So far we have considered Wilson fermions. All
this can be easily generalized to more general lattice actions, including
the theoretically popular ones satisfying the
Ginsparg-Wilson relation \cite{gw},
\bb
S^{\theta''}_{\rm GW}=a^4\sum_x \bar\psi[e^{i\theta''\gamma_5\over 2}De^{i
\theta''\gamma_5\over 2} -me^{i\theta'\gamma_5}]\psi,
\ee
where $D$ stands for the generalized lattice Dirac operator,
which does {\it not} anticommute with $\gamma_5$.
It is to be noted that the phase $\theta''$ has to be introduced through
an ordinary chiral transformation, under which $D$ is {\it not} invariant.
If $\theta''=\theta'$, this action is again invariant under the lattice
version of (\ref{parity}), so that this parity symmetry and CP are
seen to be anomaly-free. The measure does not break the
symmetry as the chiral transformation involved in (\ref{parity}) is
only a normal chiral transformation, not involving $D$: for such
transformations, the measure changes trivially, {\it i.e.,} there is
no Jacobian factor different from unity.
When $\theta''\neq\theta'$, the transformation
(\ref{parity}) alters $\theta'$ to $2\theta''-\theta'$, and the measure
changes trivially, so that the parity violating parameter is again $\theta'
-\theta''\equiv\bar\theta_f$. As before, this violation is an artefact
and has to disappear in the continuum limit.

\section{Conclusion}
The chiral phase $\theta'$ in the quark mass term cannot be held
responsible for CP violation, as demonstrated
clearly with a lattice regularization.
Lattice computations with the $\theta$ term, if possible, could be
of use in giving an indication of the strength of CP violation
resulting from this term.
Nonzero CP violation in QCD, if ever detected in experiments,
can be accommodated only through a nonzero $\theta$ term.

\section*{Acknowledgment}

It is a pleasure to thank H. Banerjee for a discussion.

\newpage
\section*{Appendix: Discussion of some  experts' comments }

{\bf ~~~~ Excerpts from comment:}

{\it ``It is not clear to me what is the question to which the paper tries
to give an answer. I thought that the concepts of anomaly and lattice
artifacts  were understood  a long  time  ago  (if  indeed  this is  the
subject is the subject {\bf [sic]}  of  the  paper).'' }

\bigskip

It would be good if the subjects were understood by everybody!

\bigskip

{\bf Excerpts from comment:}

{\it ``The action he considers is the most general one
under the standard parity but not under the new parity definition. A
mass term with a complex phase (the same which appears in the symmetry
transformation) is an eigenstate of the new parity transformation
(+1).  Therefore a general parity violating action ...
would be to add a combination of two terms
	    $\bar\psi  m e^{i\theta' \gamma_5} \psi$
and
	    $\bar\psi  m \gamma_5 e^{i\theta' \gamma_5} \psi$
without changing the phase in the Wilson term....
This is not defining a different regularization but
just adding the same content to the theory as in the
standard case.''}

\bigskip

As this expert knows, it is elementary to rewrite
this mass term in the standard form with a
change of parameters. Correspondingly, the parity symmetry
gets altered, and the phase in the Wilson term too must be altered
to preserve this symmetry.

\bigskip

{\bf Excerpts from comments on a related continuum article:}

{\it ``For a clear discussion consider a lattice regularization (as the author
did ...) where, for finite 'a', everything is explicitely defined.
As the author says, one can introduce gauge field dependent phases in
the measure in a QCD like theory. One might consider this as an extra
contribution to the action.

The action should be local, gauge invariant
and Euclidean rotation (Lorentz) invariant. These constraints allow a
nonzero theta term and also a nonzero phase in the fermion mass matrix.
Due to the anomaly, these phases are connected, and we can put all the
phases in theta.

 ...There is no CP anomaly here which would prevent us to define a CP
symmetric QCD. On that point the author is correct...''}

\bigskip

While these remarks tend to agree with this article,
a clarification is needed here. The phases are indeed ``connected''
in the continuum, if there is no
regularization and if the usual measure is used.
But ``for a clear discussion'' without going into the symmetry of the measure,
one has to remain on the lattice.
Then these phases are no longer connected
because Jacobians are trivial for ordinary
chiral rotations. The phases of the mass term
and the regulator term are of course connected, but not because
of any anomaly.

\end{document}